\documentclass[twocolumn,aps,showpacs, superscriptaddress,prl]{revtex4}
\usepackage{graphicx}
\usepackage{amssymb}
\usepackage{color}

\begin{document}

\title{Cooperativity and Frustration in Protein-Mediated Parallel Actin
Bundles}
\author{Homin Shin}
\affiliation{Department of Polymer Science and Engineering,
University of Massachusetts, Amherst, MA 01003, USA}
\author{Kirstin R. Purdy Drew}
\affiliation{Department of Materials Science and Engineering,
University of Illinois at Urbana-Champaign, Urbana, IL 61801, USA}
\author{James R. Bartles}
\affiliation{Department of Cell and Molecular Biology,
Northwestern University, Feinberg School of Medicine, Chicago, IL
60611, USA}
\author{Gerard C. L. Wong}
\affiliation{Department of Materials Science and Engineering,
University of Illinois at Urbana-Champaign, Urbana, IL 61801, USA}
\altaffiliation{Department of Bioengineering, California
NanoSystems Institute, University of California at Los Angeles,
Los Angeles, CA 90024, USA}
\author{Gregory M. Grason}
\affiliation{Department of Polymer Science and Engineering,
University of Massachusetts, Amherst, MA 01003, USA}

\begin{abstract}
We examine the mechanism of bundling of cytoskeletal actin
filaments by two representative bundling proteins, fascin and
espin. Small-angle X-ray studies show that increased binding from linkers drives a systematic \textit{overtwist} of actin filaments from their native state, which occurs
in a linker-dependent fashion. Fascin bundles actin into a
continuous spectrum of intermediate twist states, while espin only
allows for untwisted actin filaments and fully-overtwisted
bundles. Based on a coarse-grained, statistical model of protein
binding, we show that the interplay between binding geometry and
the intrinsic \textit{flexibility} of linkers mediates cooperative
binding in the bundle. We attribute the respective
continuous/discontinous bundling mechanisms of fascin/espin to
differences in the stiffness of linker bonds themselves.
\end{abstract}

\date{\today}
\maketitle

Actin binding proteins (ABP) that direct the assembly of F-actin
cytoskeletal polymers are often divided into two classes, those
that induce formation of networks, and those that induce formation
of finite-sized parallel bundles~\cite{alberts}. These motifs have
been observed for a variety of linkers, from ABP's to simple
multivalent ions, and have been
studied theoretically and experimentally~\cite%
{bausch2006,pelletier,angelini2005,wong2003,bausch2008,
wong2007,grason_07,liu1998,stevens,haviv,gov}. Espin and fascin are two
representative bundle-forming ABP's. Espins are found in
mechanosensory microvilli and microvillar derivatives, while
fascin is typically found in filopodia. Although the gross
structure of the induced F-actin bundles are similar for espin and
fascin~\cite{bausch2008,wong2007}, they behave differently, and
serve cellular functions with different requirements. Here, we aim
to explore a deeper taxonomy governing the different behaviors of
bundle-forming ABP's.

In this Letter we demonstrate that while different crosslinkers
ultimately drive parallel actin bundles to the same structural
state, the thermodynamic transition to that state depends
sensitively on linker stiffness. Monitoring the structural
evolution of bundled filaments by Small-Angle X-ray Scattering
(SAXS), we find that increasing the ratio of fascin to actin leads
to a continuous overtwisting of filaments from their native
symmetry. In contrast, crosslinking by espin produces a
coexistence of two populations, one with the fully overtwisted
geometry, and one with native twist. We propose a coarse-grained
lattice model of crosslinking in actin bundles to capture the
interplay between filament and crosslinker flexibility as well the
incommensurate geometries of actin filaments and fully crosslinked
bundles. This model reveals that stiffness of crosslinking bonds
and resistance to filament torsion sensitively control the level
of \textit{cooperativity} of crosslinking at different points
along the
filament. The mean-field thermodynamics of this model predicts: 1) a \textit{%
flexible linker} regime allows a continuous increase of crosslinks
with increased chemical potential; 2) a \textit{stiff linker}
regime exhibits a highly cooperative and discontinuous linker
binding transition; and 3) a critical-end point separating these
regimes. The respective continuous and discontinuous changes in
filament overtwist measured by scattering can be correlated with
the \textit{flexible linker} and \textit{stiff linker} regimes of
the lattice model, where a similar response to increased
crosslinking is predicted, suggesting that a small differences in
linker structure lead qualitative differences in global phase
behavior of the cytoskeleton.

\begin{figure*}[tbp]
\centering
\includegraphics[scale=0.95]{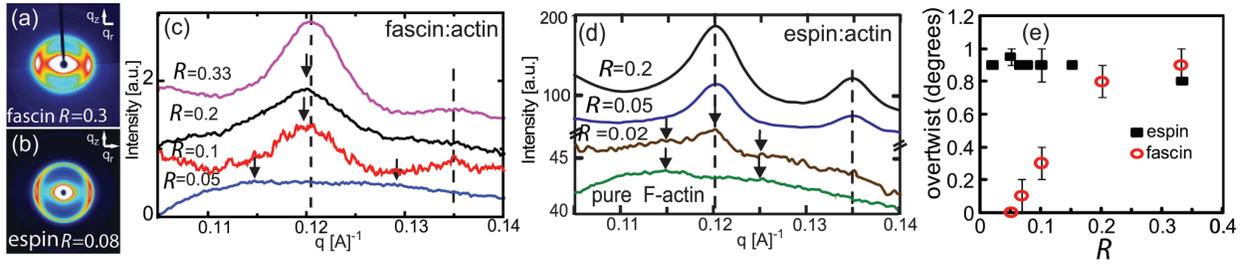}
\caption{Experimental evidence of first and second order twisting
transitions. 2D SAXS images of (a) fascin-actin bundles and (b)
espin-actin bundles. Circularly averaged SAXS data showing first
and second layer line peaks for (c) fascin-actin bundles and (d)
espin-actin bundles\protect\cite {wong2007} as a function of R.
Data is shown with a Pseudo-Voigt background subtraction.
Arrows show position of first layer line peak maximum in (c) and
the position of the unbundled layerline peaks at 0.114 and
0.125$\mathring{A}^{-1}$.  (e) Measured twist of actin bundles
as a function of $R$. } \label{fascinvespin}
\end{figure*}

To prepare X-ray samples, fresh F-actin and crosslinking protein
were mixed at specific molar ratios
$R=N_{crosslinker}/N_{G-actin}$, with 0.15 mg F-actin. F-actin was
prepared from rabbit skeletal muscle G-actin monomer
(Cytoskeleton, Inc.) which was first resuspended in 5 mM Tris, 0.2
mM $ \mathrm{CaCl_{2}}$, 0.5 mM ATP, 0.2 mM dithotreithol and
0.01\% $\mathrm{ NaN_{3}}$,  pH 8.0 and then
polymerized into F-actin by adding 100 mM KCl. F-actin was then
treated with human plasma gelsolin (Cytoskeleton, Inc.) to control
average F-actin length ($\sim 1\,\mu \mathrm{ m}$) and with
phalloidin to prevent depolymerization~\cite{janmey}. The F-actin
solution was then centrifuged at 100 000$\times $g for 1 hour to
remove polymerization buffer and resuspended in E-buffer: 0.1 M
KCl, 10 mM HEPES, 1 mM dithotreithol, 1.5 mM $\mathrm{NaN_{3}}$,
pH 7.4. Crosslinking proteins included recombinant rat espin 3A (34.3 kDa) and recombinant
human fascin (57.8 kDa), which were expressed in bacteria with an
N-terminal 6$\times $His tag, affinity purified under
non-denaturing conditions and dialyzed into E-buffer. Samples of
F-actin mixed with crosslinker were mixed, incubated, and
centrifuged in sealed quartz capillaries. SAXS experiments were
performed at 9 KeV at beam line 4-2 of the Stanford Synchrotron
Radiation Lightsource and at 12 KeV at the BESSRC-CAT (beam line
12-ID) at the Advanced Photon Source. The scattered radiation was
collected using an MAR Research CCD camera (pixel size = 79 $\mu
\mathrm{m}$). The sample-to-detector distances are set such that
the q-range is $0.01<q<0.2\,\mathrm{\mathring{A}^{-1}}$, where
$q=(4\pi \sin \theta )/\lambda $, $\lambda $ is the x-ray
wavelength, and $2\theta $ is the scattering angle. The 2D SAXS
data from both beamlines have been checked for mutual consistency.
As described previously~\cite{wong2007}, the twist of the actin
filaments when bundled with crosslinking proteins was determined
by fitting 2D SAXS data to the four sphere model of
variably-twisted F-actin convolved with the bundle structure
factor~\cite {al-khayat, angelini2005}.

The structure of espin-actin and fascin-actin bundles has been
previously
investigated~\cite{bausch2008,wong2007,tilney,bartles,derosier},
although the thermodynamic phase behavior of these actin +ABP
systems has not been mapped out. SAXS data for F-actin condensed
by fascin or espin are presented
in Fig.~\ref{fascinvespin}. The circularly averaged peak positions
of the hexagonally coordinated fascin-actin bundle are similar to
those of the espin-actin bundles. Peaks were found at 0.057,
0.100, 0.120, 0.134 A$^{-1}$ for both espin and fascin mediated
bundles at high $R$, with the first two corresponding to the
inter-actin structure factor peaks, and the latter corresponding
to intra-actin helical layer line peaks. The inter-actin spacing
for the fascin-actin bundles obtained from the position of the
$q_{10}$ peak (most intense peak visible in
Fig.~\ref{fascinvespin}a,b, was equal to $4\pi /(
\sqrt{3}q_{10})=12.9\pm 0.3\,\mathrm{nm}$, slightly larger than
that of espin-actin bundles, $12.6\pm
0.2\,\mathrm{nm}$~\cite{wong2007}. This corresponds to a fascin
size of $5.4\pm 0.3\,\mathrm{nm}$, and an espin size of $5.1\pm
0.2\,\mathrm{nm}$ using an F-actin diameter of 7.5 nm~\cite
{holmes}. Using the 4-sphere model, we found that the position of
the espin-actin bundle layer lines indicated a F-actin overtwist
of $0.9\pm 0.2$ degrees from the native left-handed $13/6$
monomers/turn twist symmetry of unbundled F-actin to a symmetry of
28/13 monomers/turn~\cite{wong2007}. Furthermore, at low $R$,
coexisting bundled and unbundled phases are observed in the
espin-actin system, as in Fig.~\ref{fascinvespin}d at $R=0.05$
where broad $13/6$ layer line peaks at 0.114 and 0.125 A$^{-1}$
can be observed simultaneously with sharp Gaussian peaks of the
overtwisted hexagonal bundles. This 2-phase coexistence in the
espin-actin bundle data, and constant layer line peak position is
in strong contrast to the small, systematic shift of the first
layer line peak of the fascin-actin bundles towards higher $q$
observed with increasing $R$ (Fig.~\ref{fascinvespin} c).This
systematic shift in peak position is only visible in the layer
line peaks of the fascin-actin system, and not in the inter-actin
structure factor peaks, indicating that it is the F-actin twist
which is gradually increasing from the native F-actin unbundled
twist symmetry ($13/6$ monomers/ turn) with increasing fascin
concentration, with a maximum of $ \sim 0.9$ degrees of over-twist
at high $R $, in agreement with recent
measurements~\cite{bausch2008,derosier}. The contrast in twisting
behavior for espin and fascin mediated bundles is summarized in
Fig.~\ref {fascinvespin}e. This fascin-actin bundle data shows a
similar decrease in twist with decreasing fascin concentration to
that previously published~\cite {bausch2008}. Espin-crosslinked actin exhibits a jump between coexisting ``low"
and ``high" twist states  with increasing espin concentration via
a first order transition, while fascin-crosslinked actin exhibits
gradual twist changes from \textquotedblleft low" to
\textquotedblleft high" twist with increasing fascin
concentration, suggesting a continuous thermodynamic pathway.

\begin{figure}[b]
\centering
\includegraphics[angle=-90, scale=0.3]{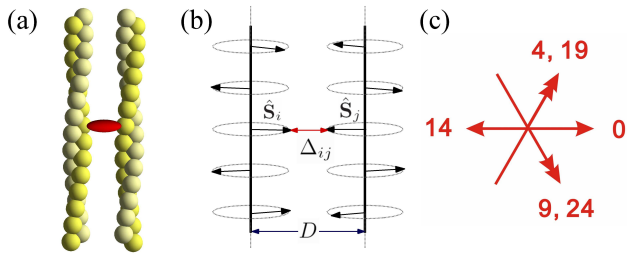}
\caption{(a) A schematic picture of two actin filaments linked by
a crosslink. (b) G-actin monomers in filaments are represented by
a set of XY spins. (c) The top view of the angular distribution of
crosslinkers (red arrows) in the unit cell of 28/13 groundstate.}
\label{actin}
\end{figure}

To study cross-linking thermodynamics, we introduce a
coarse-grained lattice model reflecting the intrinsic geometrical
frustration of parallel actin bundles. The helical axes of actin
filaments are positioned at the vertices of a hexagonal lattice
with spacing $D$. The helical configuration of G-actin monomers in
each filament is described by a set of XY ``spins" on planes
spaced along the backbone of the filament, as pictured in
Fig.~\ref {actin}. The $i$th monomer is then represented by a spin
vector $a\hat{ \mathbf{S}}_i$, where $a$ is the monomer size. The
positions described by these vectors serve as a proxy for the
locations of binding sites on the monomers themselves. For the
native configuration of actin filaments with 13/6 symmetry, the
spins wind around the filament axis by $\omega_0=12 \pi/13 $ per
monomer (see Fig.~\ref{actin}(a)).

We introduce a Hamiltonian, described by a set of spins for
filament configurations, as well as the binding site occupancy
$n_{ij} $ between two monomers, $i$ and $j$, separated by a
distance $\Delta_{ij}$,
\begin{equation}  \label{bonds}
\mathcal{H}_{binding} =\sum_{ij}n_{ij}[-\epsilon_0+\frac{k}{2}
(\Delta_{ij}-\Delta_0)^2] \ ,
\end{equation}
where the sum runs over sites on neighboring filaments and
$n_{ij}$ equals 0 or 1. Here $\epsilon_0$ describes the minimum (distortion free) energy of optimal binding and $k$ the stiffness of linkers, defining the
energy cost to deform the linkers from an aligned state,
$\Delta_{ij}=\Delta_0$. In this model, we consider the in-plane
crosslinks so that the square-deformation has a rather simple
form, $(\Delta_{ij}-\Delta_0)^2 \simeq C_0-2a^2\hat
{\mathbf{D}}_{ij} \cdot (\hat{ \mathbf{S}}_i-\hat{\mathbf{S}}_j)
+\mathcal{O}(\mathbf{S}^2)$, where $\hat { \mathbf{D}}_{ij}$ is
the unit vector of a lattice direction. Hence, protein
crosslinking occurs more favorably where two monomers co-orient
along the directions of the lattice packing. Based on this model we predict that a unique regular structure maximizes the number of \textit{perfectly aligned} crosslinks/monomer in the bundle, while requiring minimal distortion of the intrinsic twist of the filament~\cite{tobepublished}. The structure is composed of 4 sections of 5-monomers with 30/14 (overtwisted) symmetry and 2 sections of 4-monomers with 24/11 (undertwisted) symmetry, so that 6 monomers/repeat align perfectly with six-fold lattice directions and all bonds from neighboring filaments are coincident.  This structure fulfills an overall repeat unit of 28 monomers per 13 turns, consistent with the overtwisted $28/13$ geometry observed by scattering.  Based on an exhaustive search, we have found that hexagonal tilings of alternative composite structures of up 40 monomers/repeat -- including the corresponding 13/6 structure -- have smaller fraction of bound monomers than $6/28 \simeq 0.214$, provided by the composite 28/13 geometry.

The conformational adjustments of filaments required for optimal
binding give rise to \textit{cooperative} crosslinking, mediated
by torsional fluctuations of filament and linker flexibility. To
demonstrate this, we adopt a continuum model for twist
distortions, given by angular deviations from the native filament
geometry, $\mathcal{H}_0 = \frac{C}{2} \sum_\ell(\Delta
\phi_\ell-\omega_0)^2$, where $\ell$ denotes the vertical layer,
$C$ is the torsional stiffness and
$\Delta\phi_\ell=\phi_\ell-\phi_{ \ell-1}$ is the azimuthal angle
difference between two adjacent monomers \textit{along} the
filament. Based on the geometric distortion of bonds the energy
for adding a bond at a layer $\ell$ can be written as, $
-\epsilon_0+U(1-\cos[\phi_\ell-\phi_m])$. Here, $(\ell,m)$ label
the vertical and angular position of bonds, and $\phi_m =2 \pi m
/6$ indicates the preferred 6-fold direction of monomer
orientation. $U$ is a measure of linker flexibility, $U \approx
ka^2$. In our model the 28/13 groundstate packing maximizes the
number of coincident monomers from neighboring filaments, allowing
a particular large number of favorable crosslinks to form. There
are six monomers in a repeat unit of 28 monomers, which are $
(0,0);$ $(4,1);$ $(9,-1);$ $(14,3);$ $(19,1);$ and $(24,-1)$ (see
Fig.~\ref {actin}(c)).

Given a distribution of crosslinkers, we integrate out the spin
degrees of freedom via a thermodynamic perturbation theory. To
lowest order this yields an effective Hamiltonian in terms of
crosslinks alone,
\begin{equation}
\mathcal{H}_{eff}\simeq -\sum_{\ell }n_{\ell ,m}\epsilon
_{0}^{\prime }- \frac{1}{2}\sum_{\ell ,\ell ^{\prime }}n_{\ell
,m}V(\ell ,\ell ^{\prime })n_{\ell ^{\prime },m^{\prime }},
\label{H_eff}
\end{equation}
where $\epsilon _{0}^{\prime }=\epsilon_0 -U/2$. $V(\ell ,\ell
^{\prime })$ is a pairwise coupling between cross-linking of
different monomers along a filament,
\begin{equation}
V(\ell ,\ell ^{\prime })=\frac{\beta U^{2}}{2}\cos \left[ \omega
_{0}(\ell -\ell ^{\prime })-2\pi (m-m^{\prime })/6\right]
e^{-|\ell -\ell ^{\prime }|/\xi _{t}}\ .
\end{equation}
Here $\xi _{t}=2\beta C$ is the \textit{twist persistence length},
over which the orientational correlations of the native filament
geometry are \textquotedblleft washed out" by torsional
fluctuations. This van der Waals-like coupling of distinct bonds
reflects statistical correlations in crosslinking along a
filament. The rigidity of a crosslinking bond at layer $ \ell $
pins the filament in an orientation where certain nearby monomers
are close to their most favorable binding direction, so that
$V(\ell ,\ell ^{\prime })>0$. Hence, the range and strength of
$V(\ell ,\ell ^{\prime })$ are determined by $\xi _{t}$ and $U$,
respectively.

\begin{figure}[b]
\centering
\includegraphics[scale=0.3]{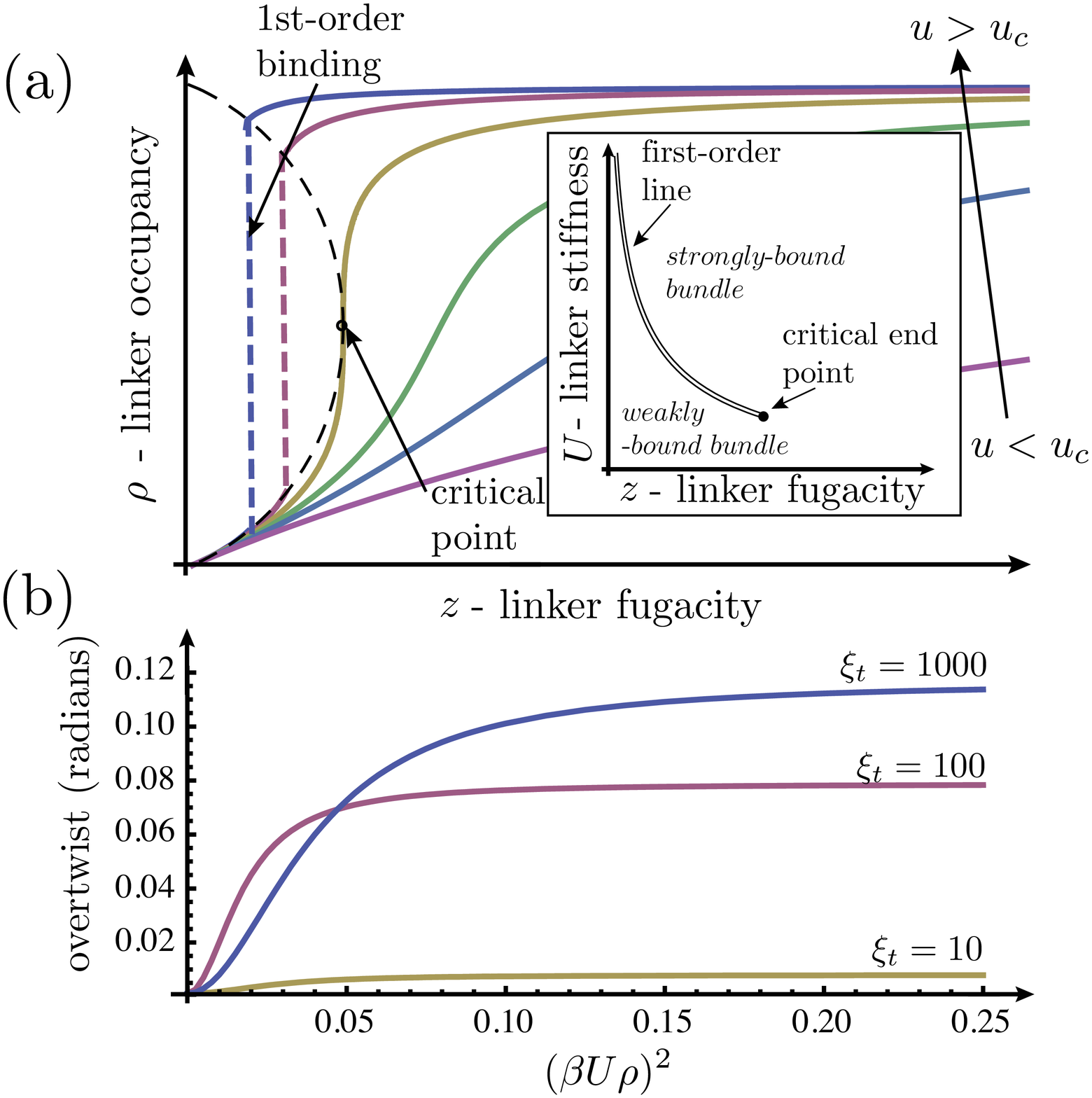}
\caption{(a)The predicted dependence of the mean site occupancy
$\protect \rho $ on linker fugacity. The inset shows the phase
diagram for fixed $z$ and $U$. (b) The correlation between
overtwist measured by $\mathrm{Im}[\ln g]$ and $\protect\rho$ for
given values of $\protect\xi_t$.} \label{rho_ot}
\end{figure}

The form of the effective binding model suggests that the
statistics crosslinker of binding falls into the Ising or
Bragg-Williams universality class. We analyze the mean-field
thermodynamics within the grand canonical ensemble at fixed
chemical potential, $\mu$, which regulates the cost of removing a
cross-linking protein from solution. Assuming a constant mean site
occupancy for the sites of the groundstate are occupied with a
probability $\langle n_{\ell,m}\rangle =\rho$, the mean-field
equation of state is determined by the solution to the
self-consistency condition, $\rho=
\big(1+z^{-1}e^{-u\rho} \big)^{-1}$. Here, $z=\exp[\beta(\mu+\epsilon^{
\prime}_0)]$ is the effective fugacity of crosslinks, proportional to the concentration of unbound linkers in solution, and $u$ is a measure of the net cooperativity of crosslinking.  Specifically, $u = N_b^{-1} \beta \sum^{\prime}_{\ell^{\prime}\not=\ell}
V(\ell,\ell^{\prime})$ where the sum is carried out over the total
$N_b$ possible sites in the 28/13 groundstate along a single
filament.  While cooperativity monotonically increases with linker
stiffness, $u \propto (\beta U)^2$, this parameter has a more
complex dependence on torsional rigidity. For small $\xi_t$,
high-temperature, cooperative binding only occurs over short
distances, so that $u \sim\xi_t$. At larger values of $\xi_t$ the
incommensurability between the native 13/6 and 28/13 twist
symmetries requires significant distortions of either the
filaments or the bonds between them. The incommensurate effects at
long range lead to a reduction of $u$ at large $\xi_t$ and maximum
value around $ \xi_t \approx 60$.

The predicted mean-field equation of state is shown in
Fig.~\ref{rho_ot}. For low cooperativity, $u<u_c=4$, $\rho$ is a
continuously increasing function of $z$, as crosslinking at
distinct sites occurs largely independently in this regime.
Increasing linker stiffness, increases the correlations in binding
events, as indicated by rise in maximum linker susceptibility,
$\chi_{\rho}=d \rho/d z$, for larger $u$. At the critical point
$u=u_c$, this susceptibility diverges at $\rho=1/2$, $\chi_{\rho}
\sim |z-z_c|^{-2/3}$, indicating a second order transition.  For
$u>u_c$, in which the stiff linkers enhances the cooperative
interactions, the transition becomes first order with a
discontinuous jump in linker density that increases with $u$.
Owing to the Ising symmetry of $\mathcal{H}_{eff}$, this model
possesses a phase diagram for fixed $z$ and $U$ reminiscent of a
``liquid-vapor" transition, in which a line of first order
transitions terminates at a critical end point (see
Fig.~\ref{rho_ot}).  Note that the value of the critical point implies a critical stiffness of order $k_c \approx k_B T / a^2$.

A second result of this analysis is one-to-one correspondence
between mean occupancy of linker sites in bundles and filament
overtwist observed in our SAXS measurements. We analyze the
following monomer-monomer correlation function, $g(\ell
_{0})\equiv \langle \exp \big\{i(\phi _{\ell _{0}+\delta \ell
}-\phi _{\ell _{0}}-\omega _{0}\delta \ell )\big\}\rangle $, where
the factor in the exponential is the excess angle between a
monomer at $\ell _{0} $ and the next monomer $\ell _{0}+\delta
\ell $ in 28-monomer packing relative to the native 13/6 twist,
that is, the mean overtwist between neighboring bonds. We
calculate $g(\ell _{0})$ perturbatively to $\mathcal{O} (U^{2})$
for the given groundstate and find overtwist, as
measured by $\mathrm{Im}[\ln g]$, to be continuously increasing
function of $ \rho $ for any given values of linker and filament
stiffness (see Fig.~\ref {rho_ot}(b)). Indeed, because neighboring
pairs of occupied bonds exert a torque on the filament to align
monomers to the groundstate symmetry, it can be shown that
$\mathrm{Im}[\ln g]\sim \rho ^{2}$ in the $U\rightarrow 0$ limit.
Hence, the continuous (discontinuous) increase in crosslinking
binding with increasing linker fugacity, implies a simultaneous
continuous (discontinuous) structural transition in terms of
filament twist.

Theoretical results here suggest that F-actin crosslinking in
parallel bundles is acutely sensitive to crosslinker flexibility.
Both predicted regimes are experimentally observed. The continuous
dependence of actin filament overtwist on the concentration of
fascin, suggests that the these linkers are too flexible to
exhibit a critical binding transition. While the comparative
insensitivity of overtwist on linker concentration in espin
bundles suggest that this binding occurs as a highly cooperative
transition, in which the rigidity of linkers immediately drives
the bundle into the fully saturated and overtwisted state. The
difference between espin and fascin binding suggests fundamental
differences in the mechanism of bundle formation (twist, diameter,
rigidity), which correlates to the distinct physiological
functions of the respective actin bundles. Hair cells require
structurally identical actin-bundles in order to mediate
reproducible mechano-chemical transduction. This may be
facilitated by an actin+crosslinker system in which the same
bundle structure is induced for a range of espin-actin molar
ratios. In contrast, fascin's function is to organize cytoskeletal
bundles in filopodial protrusions under a diverse set of
mechanical conditions~\cite{faix}, a task that may be facilitated
by the broad range of binding states and a sensitive dependence
on the fascin-actin ratio.  This view is consistent with {\it in
vivo} observations of filopodial bundles that are weakly bound by
fascin and highly dynamic~\cite{aratyn}.

\begin{acknowledgements}
HS and GMG acknowledge the support from UMass, Amherst through a
Healey Endowment Grant and thank the Aspen Center for Physics for its hospitality.  GW is supported by NSF DMR08-04363 and
the RPI-UIUC NSEC; JB is supported by NIH DC004314 and the Hugh
Knolwes Center. Part of this work is conducted at the UIUC
Frederick Seitz MRL, the Advanced Photon Source, and the Stanford
Synchrotron Radiation Lightsource.
\end{acknowledgements}

\end{document}